\documentclass[12pt,preprint]{aastex631}

\usepackage{natbib}
\usepackage{xcolor}
 
\newcommand\species[2]{#1 {\sc #2}}

\def\eg{\mbox{e.g.}}

\def\teff{\mbox{$T_{\rm eff}$}}
\def\logg{\mbox{log~{\it g}}}
\def\kmsec{\mbox{km~s$^{\rm -1}$}}

\def\vmicro{$V_{micro}$}

\shorttitle{RR Lyr Metal Line Profile Variations}
\shortauthors{Sneden et al.}

\begin{document}

\title{VARIATIONS IN LINE PROFILES OF ATOMIC TRANSITIONS IN RR LYRAE STARS}

\correspondingauthor{Christopher Sneden}
\email{chris@verdi.as.utexas.edu}

\author{Christopher Sneden}
\affiliation{Department of Astronomy and McDonald Observatory,
             The University of Texas, Austin, TX 78712, USA;
             chris@verdi.as.utexas.edu}
\author{George W. Preston}
\affiliation{Carnegie Observatories, 813 Santa Barbara Street, Pasadena,
             CA 91101, USA; gwp@obs.carnegiescience.edu}

\begin{abstract}

We have investigated the absorption shapes of atomic lines and H$\alpha$ 
in RR~Lyrae stars.
We used the database of high resolution spectra gathered with the
Las Campanas Observatory du Pont Telescope,
analyzing a set of about 2700 
short exposure spectra of 17 RRab and 5 RRc variables.
To increase the signal-to-noise of the spectra for each star, we first 
co-added spectra in small photometric phase bins, and then co-added metallic
line profiles in velocity space.
The resulting line absorption shapes vary with photometric phase in a 
consistent manner for all RRab stars, while exhibiting no obvious phase-related
variations for the RRc stars.
We interpret these line profile variations in terms of velocity gradients 
in the photospheric layers that produce absorption line profiles.
The H$\alpha$ profiles are much broader, indicative of shock temperatures of
order 100,000~K.

\end{abstract}

\section{INTRODUCTION}\label{intro}

RR~Lyrae stars are old low-mass pulsating variable stars located on the HR 
Diagram horizontal branch.
They are in the post helium flash evolutionary stage, and are undergoing
hydrostatic core helium fusion.
The relative brightness of RR~Lyraes 
absolute V-band magnitudes $<M_V>$ of +0.7)
has made them attractive
standard candles for distance determinations of Milky Way field stars, 
globular clusters and nearby galaxies.
RR~Lyraes exhibit photometric and spectroscopic variations with pulsational
periods mostly in the range 0.25$-$0.75~days, but those cyclical 
variations are essentially constant over human timescales.
Most importantly, the mean absolute V-band magnitudes of RR~Lyrae stars 
are similar with a well-calibrated dependence on metallicity.

Our group has concentrated on studies that use high resolution optical 
spectroscopy to gain insight into surface activity and envelope properties of
RR~Lyrae variables.
The goal was to explore the spectroscopic properties of these stars in
detail at many phase points with a single telescope/instrument configuration,
rather than to conduct a large-sample survey using possibly heterogeneous
data sets with limited phase information.
We have used these spectra for many purposes, 
such as determination of metallicities and abundance ratios
(\citealt{for11a,for11b}, \citealt{govea14},
\citealt{chadid17}, \citealt{sneden17}), 
studies of envelope kinematics
(\citeauthor{chadid17}, \citeauthor{sneden17}), rotation and macroturbulence
estimation \citep{preston19}, 
and description of
 helium emission/absorption variations
\citep{preston22}.
Other investigators have used these du Pont echelle spectra as bases
for studies incorporating other data sets to improve our understanding of
RR~Lyrae properties such as the Oosterhoff dichotomy 
\citep{fabrizio19,fabrizio21}, the Bailey diagram \citep{bono20}, 
the $\Delta$S metallicity parameter and [$\alpha$/Fe] ratios 
\citep{crestani21a,crestani21b}, radial velocity curves \citep{braga21}.

In our previous studies we have seen evidence of line shape variations in
atomic transitions at various pulsational phases in many RR~Lyrae stars.
Only in very limited phase domains are observed line profiles 
symmetric and described reasonably well by Gaussian functions.
In this paper we present evidence that line profile distortions are 
systematic with phase and are similar in amplitude for all stars.
These variations can be interpreted through velocity changes in photospheric 
material that is in constant motion during the pulsational periods of 
RR~Lyrae stars.
In section \S\ref{spectra} we describe the spectroscopic database 
and reduction steps for the present study.
Section \S\ref{linemeas} shows how we coadded the spectra in two ways in 
order to produce single mean profiles for each star at multiple 
pulsational phase points. 
We demonstrate that the shapes of these profiles vary systematically throughout
the pulsational cycles in essentially the same manner in all of our program
stars.
Finally In \S\ref{skewinterp} we develop a simple model for velocity 
gradients that can account for the RRab line profile variations.

\vspace*{0.2in}
\section{THE SPECTROSCOPIC DATA SET}\label{spectra}

From 2006 to 2014 we gathered more than 5400 RR~Lyrae spectra with the Las
Campanas Observatory du Pont Telescope Echelle Spectrograph\footnote{
https://www.lco.cl/technical-documentation/echelle-spectrograph-users-manual/}.
In this work we use the database of several thousand 
du Pont echelle spectra of RR~Lyrae stars \citep{sneden21b}.
The spectra have resolving power $R \equiv \lambda/\delta\lambda$ =~27,000 
at 5000~\AA.
Here we have selected a set of stars that have a large number of
spectra over their complete pulsational cycles.
In Table~\ref{tab-stars} we give the photometric properties 
of the chosen stars, along with their metallicities, the total number of 
du Pont spectra available, and the number of co-added spectra (discussed below).
The $P$ and $T_0$ values are taken from Table~1 of \cite{chadid17}.
The metallicities are means of the [\species{Fe}{i}/H] and 
[\species{Fe}{ii}/H] values listed in Table~2 of that paper.

The pulsational periods of RR Lyrae stars are mostly in the range 
0.25~d $<$ P $<$ 0.75~d.  
Spectroscopic exposure times were confined to small fractions of 
these periods, never more than 600~s ($\sim$0.01P), usually 400~s or less.
This restriction resulted in relatively poor signal-to-noise ratios, 
S/N~$\sim$~15$-$20 near 4300~\AA\ for most spectra.

The reduction procedures for the du Pont data have been described 
in detail by \cite{for11a}; see that paper for details of the methods.
Standard IRAF\footnote{
https://iraf.noirlab.edu/} 
\citep{tody86,tody93,fitzpatrick24} procedures were employed in
these tasks.
The resulting fully reduced ``raw spectra'' have about 50 echelle orders 
spanning the approximate wavelength range 3900$-$8800~\AA.
Then, for the present work we created continuous flattened spectra from
4000$-$4600~\AA, a spectral region that has the largest collection of atomic
transitions available for study in typical RR~Lyrae stars.
The wavelength step size in these raw spectra is 
$\delta\lambda$~=~0.045~\AA/pixel, or
velocity displacement $\delta V$~=~3.1~\kmsec\ at 4300~\AA.

\begin{figure}
\epsscale{0.60}
\plotone{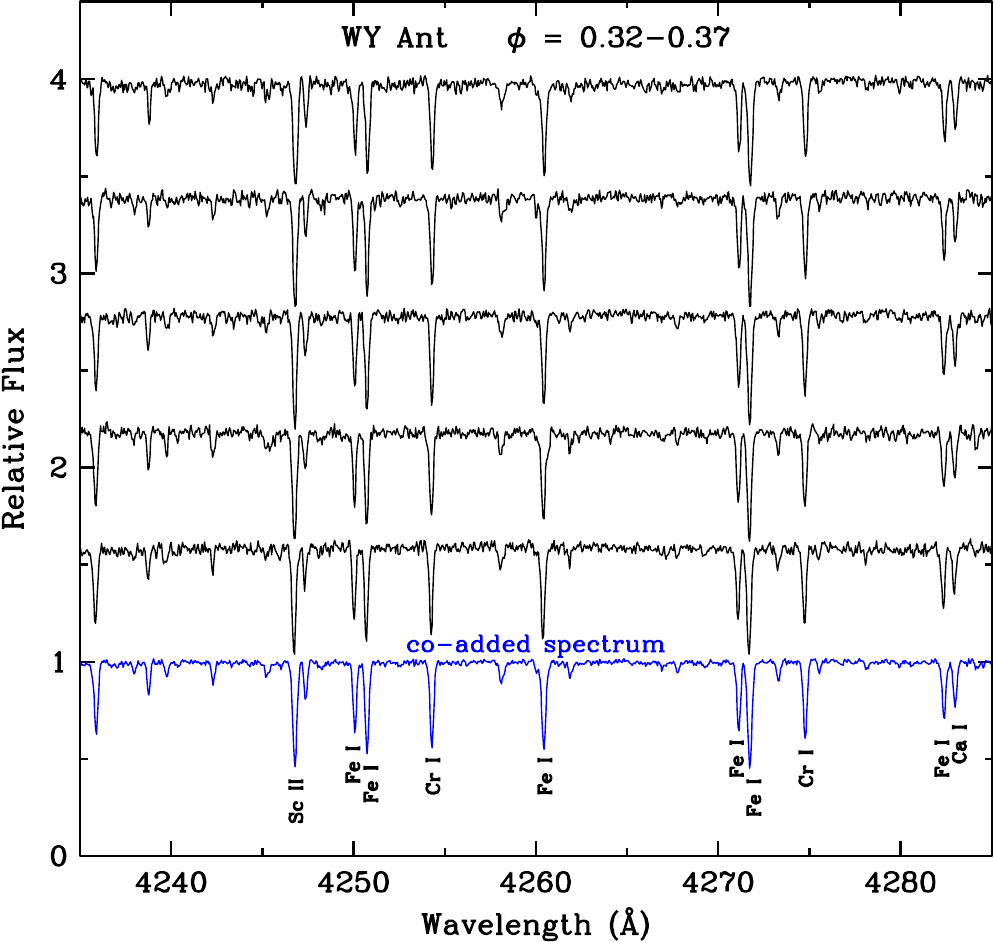}
\caption{\label{fig01}
\footnotesize
   An example of raw and co-added spectra in one spectral region at a
   single phase for program star WY~Ant.
   This relatively bright star already with good $S/N$ has been chosen for 
   display.
   The identified atomic lines are those that were measured in this
   particular co-added spectrum.
}
\end{figure} 

The observed line profiles of photospheric transitions in all stars
arise from the combined broadening effects of the spectrograph instrumental
profile and stellar temperature, microturbulence, macroturbulence, and
rotational conditions.
RR~Lyraes have continuous envelope and atmospheric motions that are
accompanied by shock waves at multiple phase points.
These motions have been well documented, and are the sources of the
asymmetries under investigation here.
\cite{preston19} attempted to detect rotation in RRab stars, concluding that
its effects cannot be disentangled from those arising from macroturbulence.
The combined effects of macroturbulence and rotation were estimated as
$V_{macrot}$~$\geq$ 5~$\pm$~1~\kmsec; see their Figure~16.

Here we follow the discussion in \cite{preston19} to estimate various
broadening components.
The instrumental broadening of the du Pont echelle was explored extensively
in that paper, which derives $<$FWHM$_{inst}>$~=~11.4~$\pm$~0.5~\kmsec.
The combined effects of thermal and microturbulent broadening,
expressed as $\sigma_{th,micro} = 2\sqrt{2kT/M + V_{micro}^2}$ can be reasonably estimated
because of the regularity of RR~Lyrae variations.             
Figure~6 of \citeauthor{preston19} suggests that for RRab stars \teff\      
varies from $\sim$6100~K in the phase range $\phi$~$\sim$0.35$-$0.75,       
to $\sim$7300~K near $\phi$~$\sim$~1.0 .                      
Similarly, microturbulent velocity
\vmicro\ changes in the range $\sim$2.9$-$4.0~\kmsec\ in concert  
with the \teff\ variations.                                   
This yields a small range of thermal and microturbulent broadening during   
RRab cycles:  3.2~$\lesssim$~$\sigma_{th,micro}$~$\lesssim$~4.2~\kmsec.     
For our purposes, understanding the imprecision of these estimates, we      
will hereafter assume $\sigma_{th,micro}$~=~3.7~\kmsec.

As a preliminary to our investigation of skew we coadded observations in 
small intervals of phase to increase the S/N.  
Several considerations are involved in the creation of coadded spectra for 
a given star, the most important being brightness of the star, its location 
in the sky, and the dates of observing runs when it is observable.  
Thus, many observations were made of stars in the 
constellations Apus and Octans
 with declinations $<$ -80$^{\circ}$, because such stars can be 
observed at very large hour angles up to $\pm$6~h. 
Conversely, limited numbers of observations were made of stars with northerly 
declinations.

The maximum number of spectra in coadds is $\sim$10 and a typical number is 5. 
In a few instances single observations are included as ``coadds'' at 
interesting phases.  
Typical phase intervals of coadds is 0.03P. 
Maximum interval with few exceptions is 0.1P.
In Figure~\ref{fig01} we illustrate the co-addition process.  
For this and some subsequent figures we have chosen program star
WY~Ant to illustrate the analytical steps.
Five raw spectra contributed to the mean spectrum shown in the figure.
Visual inspection will confirm that while the individual spectra have
$<$S/N$>$~$\sim$~50, the co-added spectrum has $<$S/N$>$~$\sim$~100, revealing
atomic lines worthy of further analysis while depressing many noise features.

\section{LINE COADDITION AND PROFILE MEASUREMENTS}\label{linemeas}

\subsection{Formation of a Mean Metallic Line}\label{linemean}

\begin{figure}
\epsscale{0.60}
\plotone{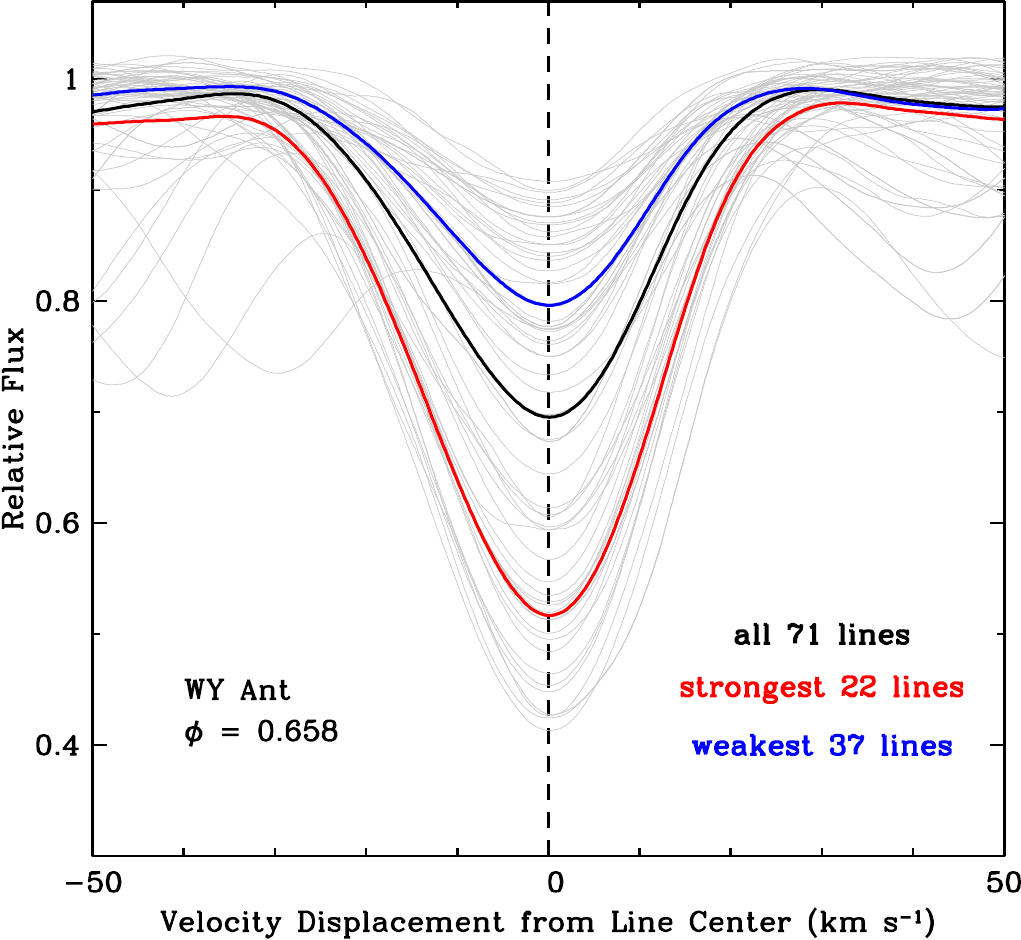}
\caption{\label{fig02}
\footnotesize
   An example of the transition co-addition process.
   The 71 individual lines used for WY~Ant at pulsational phase $\phi$~=~0.658
   are shown as faint gray profiles.
   The mean profile generated by co-addition of all of them is shown in black.
   The mean of just the weakest 37 lines is in blue, and that of the
   strongest 22 strongest lines is in red.
}
\end{figure}

\begin{figure}
\epsscale{0.60}
\plotone{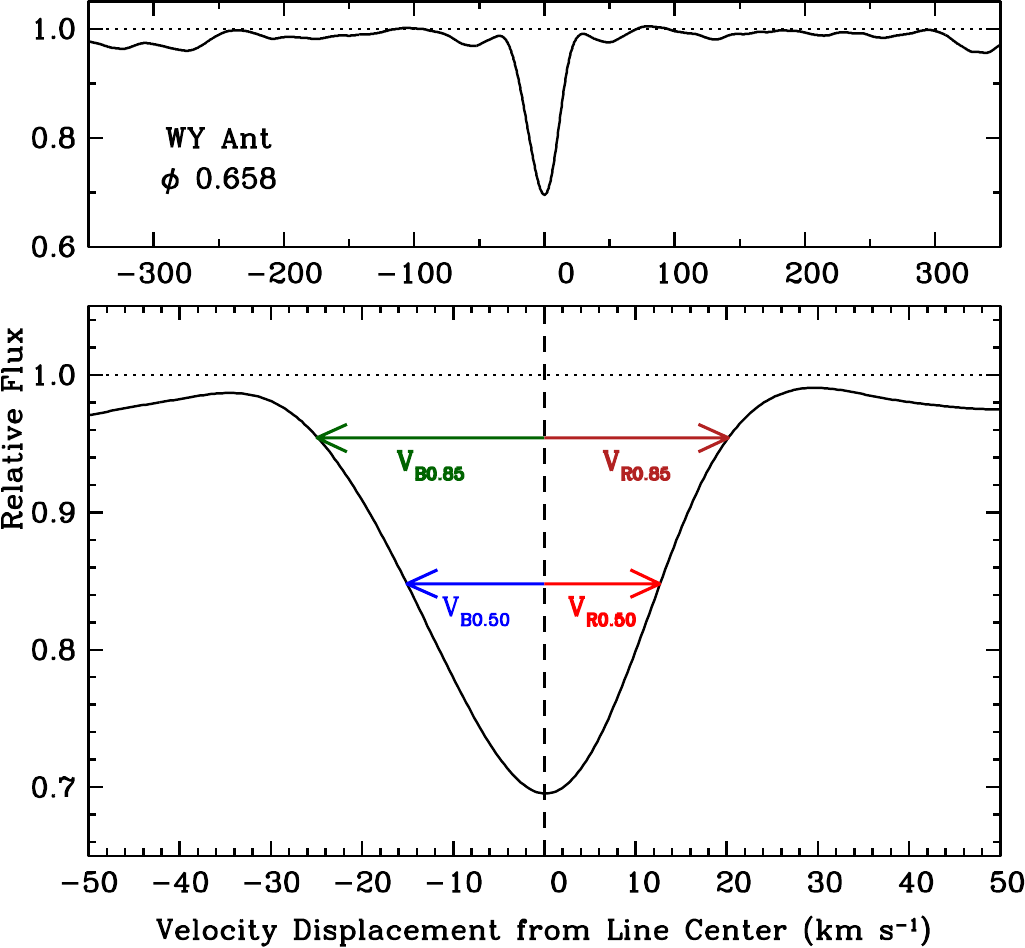}
\caption{\label{fig03}
\footnotesize
   An example of the transition measurement process.
   The displayed line profile is the co-added spectrum in velocity space,
   using all transitions that were previously shown in Figure~\ref{fig02}.
   The top panel shows a spectral region $\pm$350~\kmsec\ ($\pm$5.0~\AA\
   at 4300~\AA) surrounding the co-added transition.
   The dotted line indicates the adopted continuum.
   The bottom panel has a blowup of just the region $\pm$50~\kmsec\
   around the transition.
   The dashed line is placed at line center.
   The observed velocity widths to the blue and red at flux levels 0.50 and 
   0.85 of the continuum are indicated by arrows of different colors and
   by labels (\eg, B0.50 means the width of the blue part of line profile
   at flux level 0.50 of the continuum.
}
\end{figure}

\begin{figure}
\epsscale{0.60}
\plotone{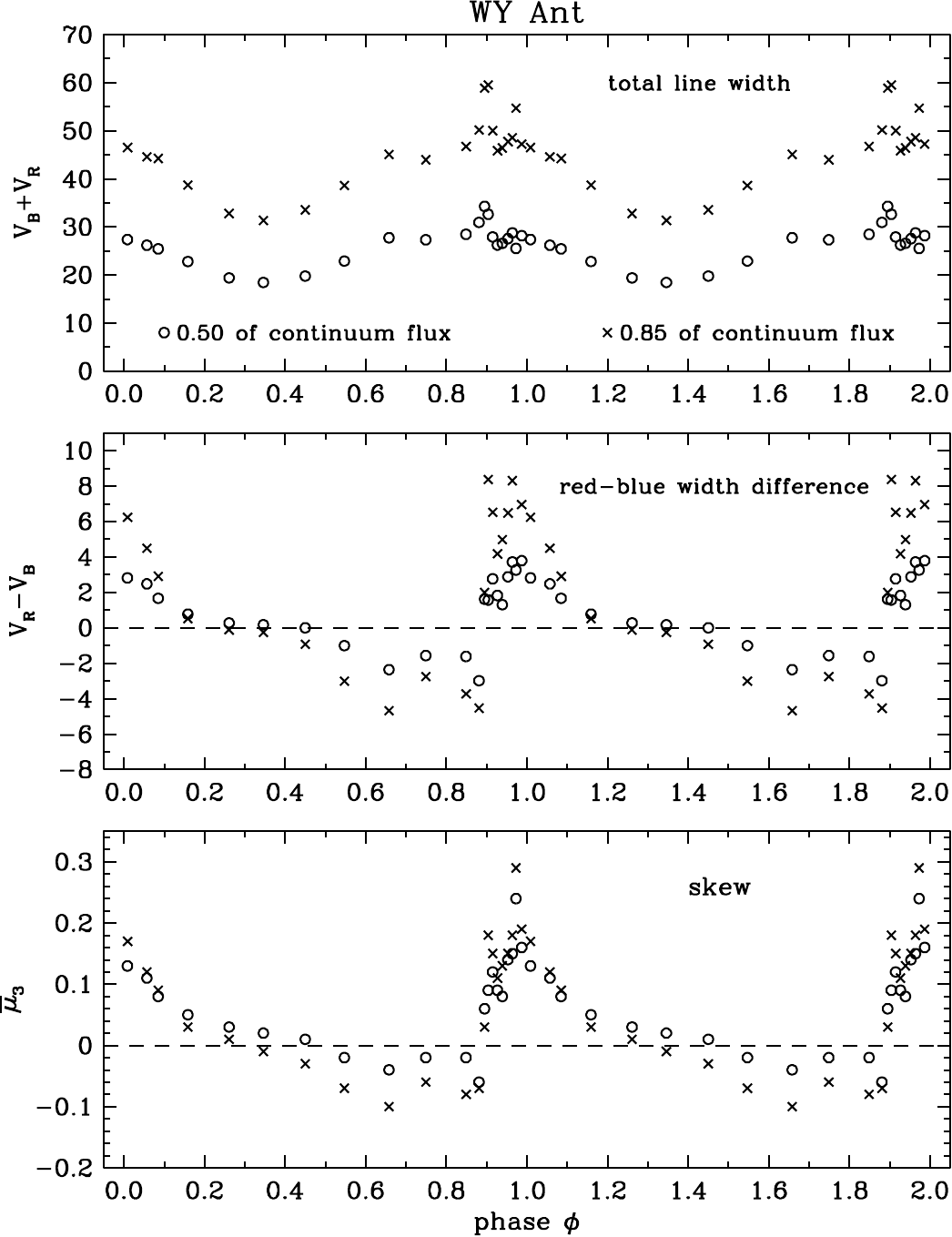}
\caption{\label{fig04}
\footnotesize
   Statistical measures for WY~Ant lines at all co-added phase points.
   In the top panel the line widths V$_{B}$+V$_{R}$ are shown.
   Open circles are used for the widths at 0.50 relative line flux level
   and $\times$ symbols for the 0.85 level, as given in the panel legend.
   The middle panel displays differences between blue and red widths 
   V$_{R}$$-$V$_{B}$, and the bottom panel contains the line skew values.
   Data for this figure are listed in Table~\ref{tab-broaden}.
}
\end{figure}

While atmospheric conditions of RR~Lyraes are constantly variable,
inspection of Figure~6 in \cite{preston19} and other literature studies 
suggest that typically for RRab stars away from the shock-wave phases 
$<$\teff$>$~$\sim$~6200~K, 
$<$\logg$>$~$\sim$~2.0, and $<$\vmicro$>$~$\sim$~3.0~\kmsec.
Superficially, their spectra look similar to that of the Sun.  
However, most RR~Lyraes are metal-poor, [Fe/H]~$<$~$-$1, and many weak lines
in the solar spectrum disappear into RR~Lyrae continuum regions.
Additionally, RR~Lyraes have much broader lines than does the Sun,
eliminating some otherwise promising lines that become inaccessible for 
transition profile analysis.
The relatively low $S/N$ values of our spectra also mangle some profiles
of the remaining lines.
Finally, the number of measurable transitions varies systematically with phase.
The sharpest and cleanest line profiles occur in the 
$\phi$~$\simeq$ 0.35~$\pm$0.10 domain, while the broadest, most distorted 
profiles are often seen when $\phi$~$\simeq$ 0.90~$\pm$0.15.

To systematically determine a candidate line list that might serve for most
program stars at most phases, we first used the solar line compendium of 
\cite{moore66} in conjunction with one of our higher $S/N$ spectra to 
identify promising transitions in the 4000$-$4600~\AA\ that should be examined
in all spectra.
This exercise yielded about 100 lines.

For each co-added spectrum, we systematically measured approximate equivalent
widths ($EW$s) in a semi-automatic procedure using the specialized software
code SPECTRE \citep{fitzpatrick87}.
The purpose was not to determine accurate $EW$s for detailed study.
Rather we wanted to identify which lines should be included in our line profile
analysis of each spectrum, and their general absorption strengths.
Figure~\ref{fig01} identifies lines retained for the example star/phase
in the spectral region displayed in the plot.

For a given spectrum, the spectral region surrounding each identified line was 
interpolated in velocity units to small steps (0.3~\kmsec). 
The central depth of the observed line was found in the interpolated spectrum,
and a velocity array extending $\sim\pm$400~\kmsec from this central point
was computed.
Then these line arrays in velocity space were co-added to yield the final
mean line profile.
The number of individual transitions contributing to the mean profiles varied
greatly from star to star (due to metallicity differences) and phase to phase 
(due to changes in line profile sharpness).
Additionally, $S/N$ differences for individual stars based on their apparent
magnitudes affected the number of lines available for analysis.
In total, the mean line profiles were composed of between $\simeq$20$-$80
transitions.

When the number of transitions used for a star at a particular phase was
large, $\gtrsim$40, the mean profiles could be constructed with subsets
of lines segregated in line strength, ionization state, and excitation
energy. 
In Figure~\ref{fig02} we show a typical example of such flexible line 
averaging, choosing WY~Ant at $\phi$~=~0.658.
We have selected this star for display in this and several more figures
because it is typical of our RRab sample, and has spectra well covering its
pulsational phases.
At $\phi$~=~0.66 WY~Ant has a co-added mean spectrum with 71 measurable 
lines having an $EW$ range $\simeq$20$-$280~m\AA.
Here we show the 71 line profiles individually as thin black lines and
their mean as a thick black line.
The means of the weakest and strongest lines are shown as blue and red lines,
respectively.
Inspection of these transition profiles reveals line asymmetries in all
three means, with extended blue wings in each case. 
These asymmetries will be quantified in \S\ref{linestats}.
In columns 9 and 10 of Table~\ref{tab-stars} we list the minimum and maximum
number of lines that participated in the line profile calculations at 
individual phase points for each star.

\subsection{Line Statistics}\label{linestats}

We made 3 sets of line profile measurements:
\textit{(1)} blue and red velocity widths, labeled $V_{B0.50}$ and 
$V_{R0.50}$, at fractional line fluxes of 0.50 of the local continuum 
(\eg, at the half widths); 
\textit{(2)} velocity widths $V_{B0.85}$ and $V_{R0.85}$ at 0.85 of the 
continuum;
and \textit{(3)} Pearson's statistical 3$^{rd}$ moment skew values\footnote{
https://en.wikipedia.org/wiki/Skewness}
at those two flux levels, $\bar{\mu}_{3,0.50}$ and $\bar{\mu}_{3,0.85}$.
In Figure~\ref{fig03} we use WY~Ant at pulsational phase $\phi$~=~0.658
to illustrate our line profile parameters.
The line profile is the one formed from the whole transition set depicted
in Figure~\ref{fig02}.
The choice of 0.50 central depth is a standard one for line profile 
descriptions.
The 0.85 level was empirically chosen to measure the line asymmetries
in the wings of the profile, without going so far from line center that
continuum placement and contamination from other spectral features become
major measurement uncertainty sources.

The first task in this process was continuum setting.
In Figure~\ref{fig02} the apparent features that are $\pm$30$-$50~\kmsec\ 
away from line center in the coadded line profile arise from atomic blends 
in individual transitions that depress their local continua.
Therefore for the coadded line profiles better continuum placement was 
accomplished from inspection of a series
of high-flux points in an expanded velocity interval, as we show in
the top panel of Figure~\ref{fig03}.
Then we measured line quantities defined in the previous paragraph;
see the bottom panel of this figure.
Their values in all program stars at all coadded phase points are listed
in Table~\ref{tab-broaden}.

In the WY~Ant $\phi$~=~0.658 example, our measured quantities were
$V_{B0.50}$~=~15.1~\kmsec,  $V_{R0.50}$~=~12.7~\kmsec,
$V_{B0.85}$~=~24.9~\kmsec, $V_{R.85}$~=~20.2~\kmsec, and skew$_{0.50}$~=~0.04,
and skew$_{0.85}$~=~$-$0.10..
The blueward asymmetry is well defined at both measurement depths:
$\delta V_{B.50}$~$\equiv$~$V_{B.50}$~+~$V_{R.50}$ = $-$2.4~\kmsec\ and
$\delta V_{B.15}$~$\equiv$~$V_{B.15}$~+~$V_{R.15}$ = $-$4.7~\kmsec.
The skew values are $\bar{\mu}_{3,0.50}$~=~$-$0.04 and 
$\bar{\mu}_{3,0.85}$~=~$-$0.10.  
These line statistics all support the visual impression in 
Figure~\ref{fig03} of a line with a blueward distorted profile.

\begin{figure}
\epsscale{0.60}
\plotone{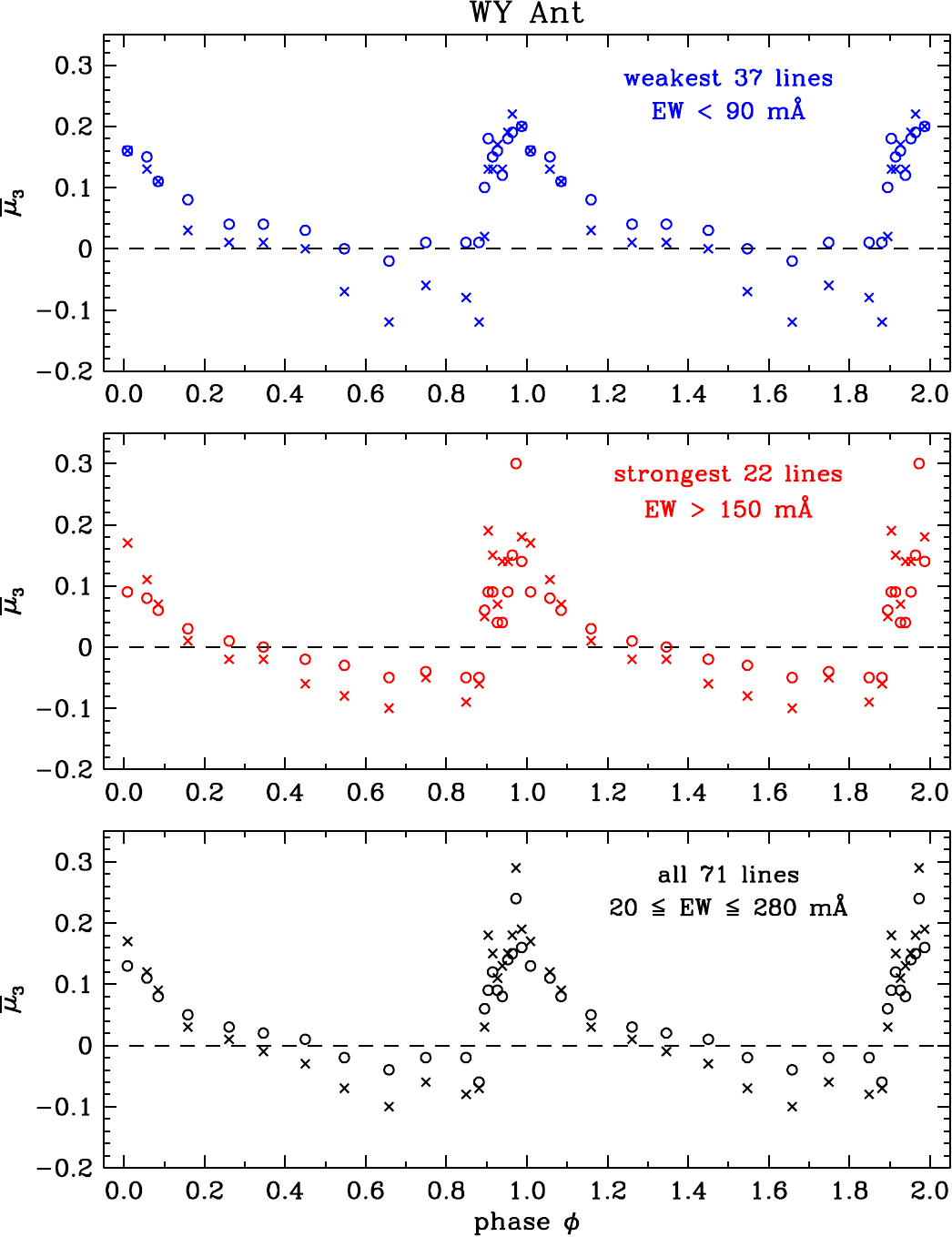}
\caption{\label{fig05}
\footnotesize
   Variations of metal line skew values for WY~Ant for weaker lines (top
   panel), strongest lines (middle panel), and the total line set (bottom
   panel).
   The bottom panel is a repeat of the bottom panel in Figure~\ref{fig04}.
}
\end{figure}

In Figure~\ref{fig04} we show three descriptors of WY~Ant line 
shapes at the 0.50 and 0.85 relative flux levels for all 21 of its co-added 
phase points.
In the top panel total line widths $V_{B0.50}$+$V_{R0.50}$ and 
$V_{B0.85}$+$V_{R0.85}$ are plotted.
A relatively smooth variation from minimum width near phase 
$\phi$~$\simeq$~0.35 to maximum near $\phi$~$\simeq$~0.90 is easy to see and
is not surprising.
This trend has been noted in many papers, and has been explored in detail by
\cite{for11b} and \cite{preston19}.
In the middle panel we show a simple asymmetry measure of the differences 
between red and blue widths V$_{R0.50}$$-$V$_{B0.50}$ and 
V$_{R0.85}$$-$V$_{B0.85}$.
At phases near line width minimum, 0.2~$\lesssim$~$\phi$~$\lesssim$~0.5,
WY~Ant metal line profiles are symmetric to our measurement limits.
This apparently calm phase interval is followed by a small but easily
detected blueward asymmetry that lasts until $\phi$~$\simeq$~0.85,
and then by a sharp turn to a much larger redward asymmetry during the
photometric rising-light phase domain.
In the bottom panel of Figure~\ref{fig04} we plot the 3$^{rd}$ moment 
skew values at the two line flux points.  
The skews are consistent with the line width differences and are more
statistically robust.
We adopt skew as the primary line asymmetry indicator.

WY Ant at phase $\phi$~=~0.658 has many useful metallic lines with a large
absorption strength range, 20~$\leq$~EW~$\leq$~280~m\AA\
(reduced width $-$5.3~$\leq$~log$_{10}$(EW/$\lambda)$~$\leq$~$-$4.2).
Following Figure~\ref{fig02}'s line strength divisions,
in Figure~\ref{fig05} we show the variation of skew for weak, strong,
and all lines.
Inspection of these figure panels strongly suggest that the skews are 
qualitatively the same for lines on the linear to the damping portions
of the curve of growth.
This further implies that the metallic line formation photospheric layer
is relatively compact. 
All atomic absorptions occurs in similar temperature/pressure conditions.

\begin{figure}
\epsscale{0.60}
\plotone{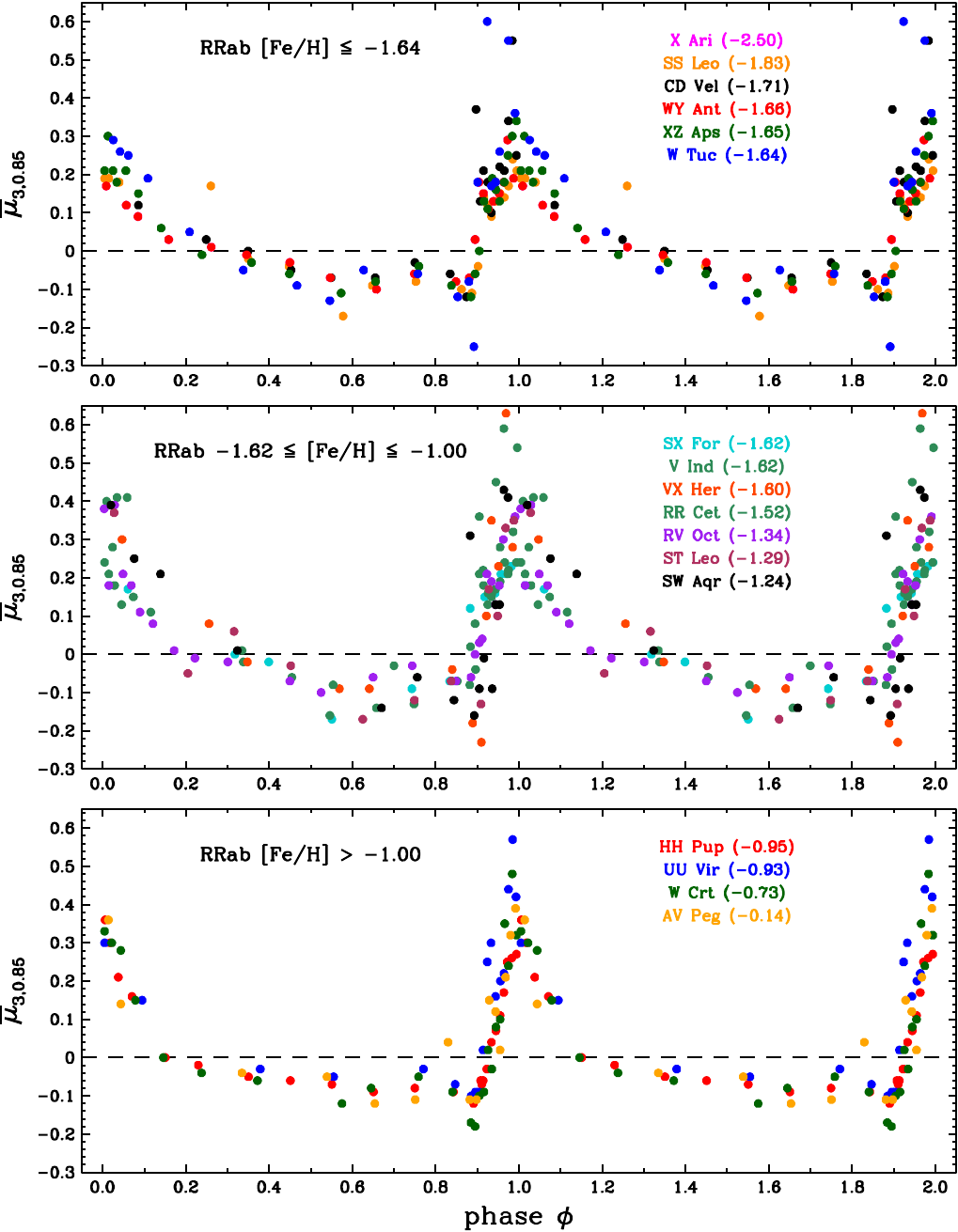}
\caption{\label{fig06}
\footnotesize
   Skew values for our sample of RRab stars, plotted as functions of phase.
   The three panels contain stars grouped for visual clarity according to 
   their metallicities, with the most metal-poor in the top panel and
   the most metal-rich in the bottom panel.
   Skew values for individual stars are indicated with different colors,
   and their metallicities from Table~\ref{tab-stars} are in the panel
   legends.
}
\end{figure}

\begin{figure}
\epsscale{0.60}
\plotone{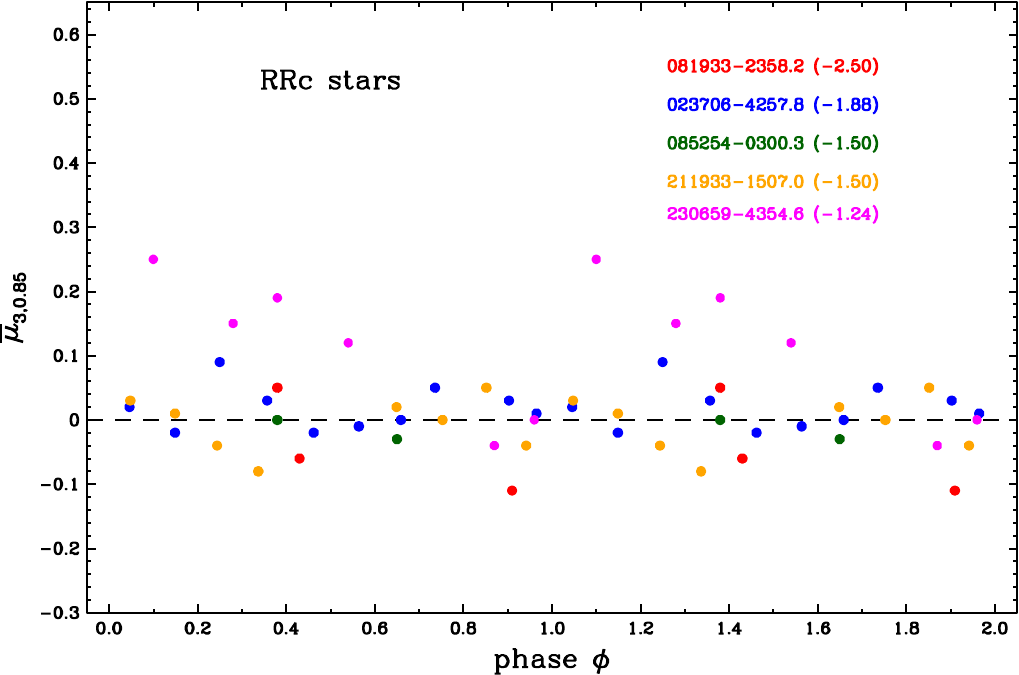}
\caption{\label{fig07}
\footnotesize
   Skew measurements for the RRc stars.
}
\end{figure}

\subsection{Line Profile Variations for RRab and RRc Stars}\label{skewabc}

Our program sample is composed of 17 RRab and 5 RRc stars.
In Figure~\ref{fig06} we gather the skew values for the RRab stars.  
Three panels are used separate the stars into lowest, intermediate,
and highest metallicity groups.
Previous papers by our group (\eg, 
\citealt{chadid17,sneden17,sneden18,preston19} have divided the RR Lyraes
into metal-rich (MR) or metal-poor (MP) categories with the split estimated
to be [Fe/H]~=~$-$1.0.
By that criterion, the top and middle panels of Figure~\ref{fig06} contain
all MP stars and the bottom panel has only MR stars.  
Inspection of the figure reveals no obvious skew differences in different
metallicity bins.
For MR stars (bottom panel) the phase regime of positive skew appears to be
slightly smaller than those of the more metal poor stars (middle and top
panels), but our MR sample has only 4 stars. 
Additionally MR stars have stronger line spectra at all phases, yielding
more sharply defined average line profiles than for the MP groups.
Gathering spectra for more MR RRab stars will be needed to resolve this 
possible difference between the RRab metallicity groups.

In Figure~\ref{fig07} we show the variation in skew with phase for 
our sample of RRc stars.
The data in this plot strongly suggest that RRc metal lines have symmetric 
profiles.  
Apart from star AS230659–4354.6, the measured skew values all are contained
within $-$0.1~$\lesssim$~$\bar{\mu}_{3,0.85}$~$\lesssim$~+0.1, and there
is no obvious phase dependence of the skew values.
We caution that we did not observe RRc stars as intensively as the RRab stars.
We have fewer coadded phase points for any RRc star (N$_{spec,max}$ = 10,
Table~\ref{tab-stars})
than for any RRab (N$_{spec,min}$ = 13, Table~\ref{tab-stars}.
Additionally, RRc line strengths are weaker than those of RRab stars of
similar metallicities because RRc stars are warmer.
Finally, the S/N values of our RRc spectra are typically lower
than those of our RRab sample.
Therefore we regard the skew range in Figure~\ref{fig07} as consistent 
with no line profile red/blue distortions.
However, this conclusion is based only on the 5 RRc stars considered
in this paper, and may not apply to all members of the RRc class.
For example, \cite{bono20b} applied hydrodynamical modeling to detailed
light curves of the RRc variable U~Com (P~=~0.29~d), finding photometric
features that would suggest that structure in RV variations might be
present.
A larger spectroscopic survey of RRc stars should be undertaken.

\section{INTERPRETATION OF THE LINE SKEW MEASUREMENTS}\label{skewinterp}

Detailed dynamical model atmosphere 
non-local thermodynamic equilibrium (NLTE) line transfer computations 
are beyond the scope of the present observational spectroscopic contribution.
Here we sketch the outlines of a possible interpretation of the 
phase-dependent skewed line shape profiles of RRab stars.

\subsection{Skew due to Putative Velocity Gradients in Metallic Line-forming 
Regions}\label{skewvgrad}

\begin{figure}
\epsscale{0.60}
\plotone{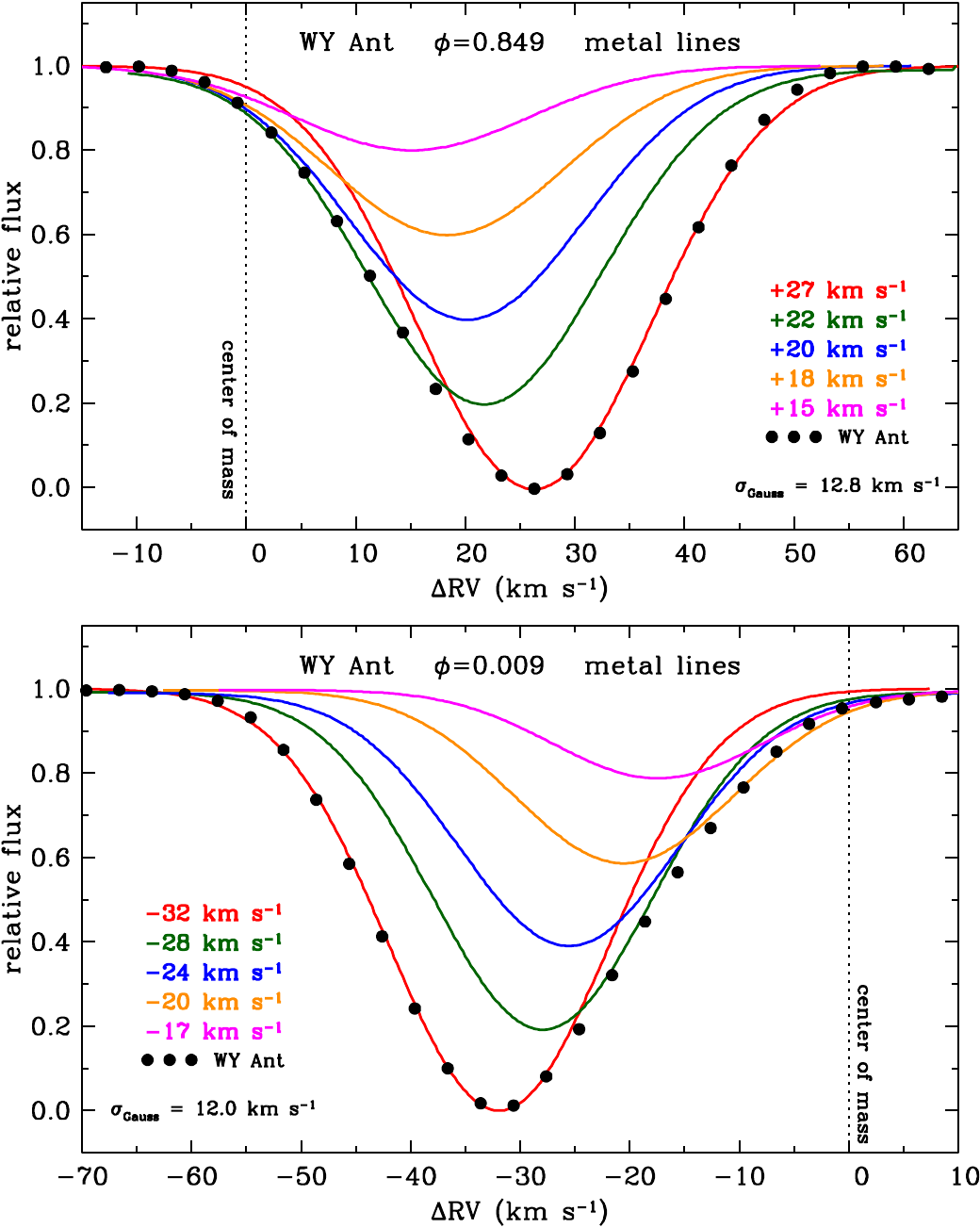}
\caption{\label{fig08}
\footnotesize
   Coadded metallic line profiles (black dots) and modeled line
   profiles (colored lines) for WY~Ant at two pulsational phases.
   The velocities $\Delta$RV are with respect to the stellar center of mass
   velocity.
   See the text for description of the empirical model line profiles; their
   relative velocities are noted in the figure legend.
}
\end{figure}

\begin{figure}
\epsscale{0.60}
\plotone{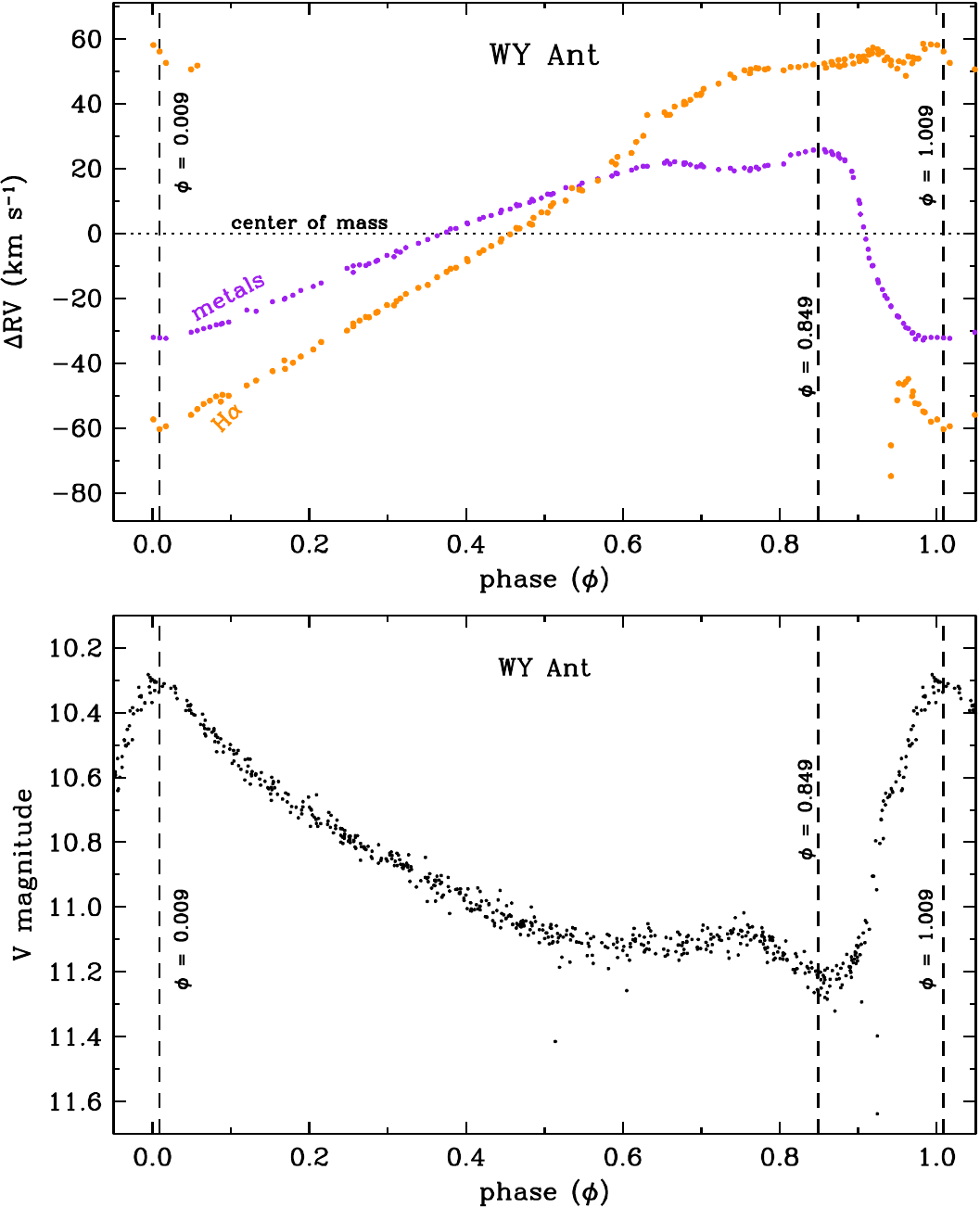}
\caption{\label{fig09}
\footnotesize
   Top panel: radial velocity variations with phase for metal lines 
   (purple color) and H$\alpha$ (orange) in WY~Ant.
   The ordinate $\Delta$RV values are measured relative to the WY~Ant
   center-of-mass velocity, 203.7~\kmsec.
   Bottom panel: Photometric V-band magnitudes for WY~Ant from the ASAS
   database (\citealt{pojmanski05} and references therein).
   In both panels vertical dashed lines denote phases $\phi$~=~0.009 
   (repeated at $\phi$~=~1.009) and 0.849 that have
   been chosen for line profile display in Figure~\ref{fig08}.
}
\end{figure}

\begin{figure}
\epsscale{0.60}
\plotone{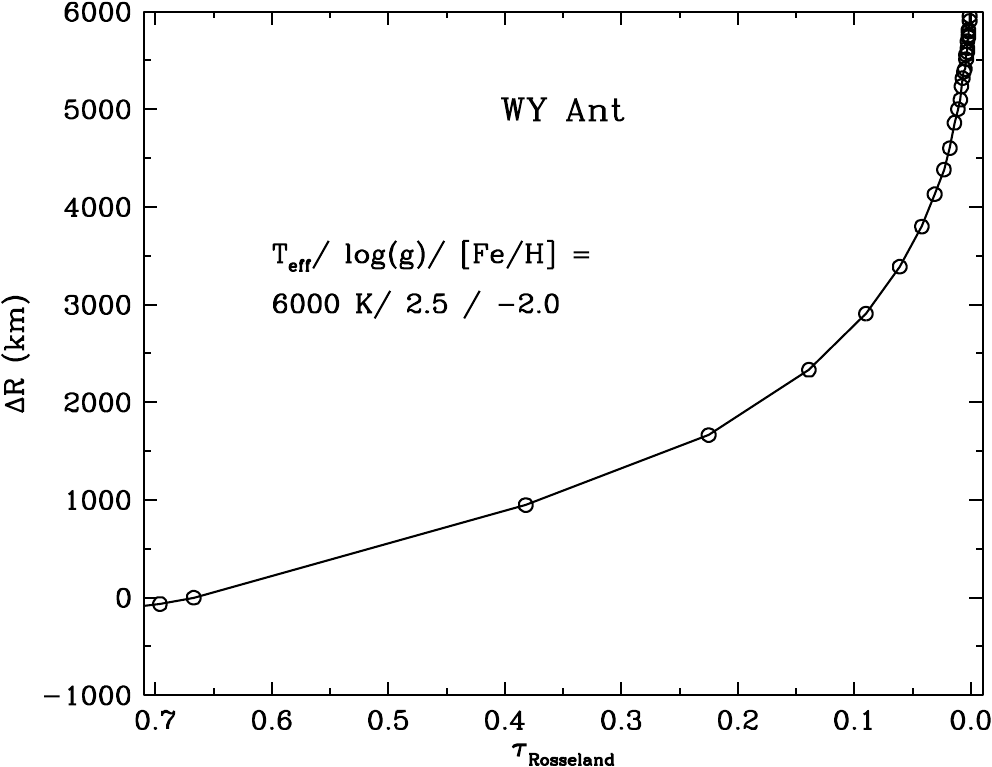}
\caption{\label{fig10}
\footnotesize
   Heights ($\Delta$R in km) 
   above Rosseland $\tau$~=~0.67 in a model from the
   \cite{kurucz11,kurucz18} grid with parameters indicated in the legend.
}
\end{figure}

Investigations of skew in solar absorption lines were reviewed in detail 
by Dravins (1982 \nocite{dravins82} and references therein), who interpreted 
this skew successfully in terms of a solar surface covered by myriads of 
hotter upflowing convection cells surrounded by cooler downflowing 
intercellular gas.
We propose a different explanation for the time-dependent skew observed in 
RR~Lyrae stars: namely, the existence of velocity gradients 
in their stellar atmospheres first predicted by \cite{bono94}.

In the top panel of Figure~\ref{fig08} we show the metallic line 
skew profiles in metal-poor WY~Ant at pulsational phase $\phi$~=~0.849, near 
minimum light (maximum infall velocity).
In the bottom panel we show the skewed line profiles at $\phi$~=~0.009,
just after maximum light (near maximum outflow velocity).  
Black filled circles denote the observed profiles, and solid lines represent
our fits to the observations, to be explained below.
The velocities, labeled $\Delta$RV, are relative to the center-of-mass velocity,
defined here to be the average of the radial velocities of metal lines
over a pulsational cycle.
For WY~Ant, we derived center-of-mass~RV~=~203.7~\kmsec, using velocity data
originally presented in \cite{chadid17}.
The $\Delta$RV values for both metal lines and H$\alpha$ are shown in 
the top panel of Figure~\ref{fig09}, with vertical lines placed at the 
phases chosen for Figure~\ref{fig08}.

The bottom panel of Figure~\ref{fig09} contains the photometric 
V-band light curve of WY~Ant\footnote{V magnitudes were taken from the All Sky 
Automated Survey (ASAS, \citealt{pojmanski05}); and references therein;
https://www.astrouw.edu.pl/asas/}.
This connects photometric and spectroscopic quantities for WY~Ant.
Other RRab stars in our study have essentially the same variations throughout
their pulsational cycles.
Thus, phase $\phi$~=~0.849 shows very blue-skewed line profiles 
(Figure~\ref{fig05}) for WY~Ant, accompanying maximum infall velocity 
($\Delta$RV~$\simeq$~25~\kmsec; Figure~\ref{fig09} top panel) and
minimum brightness ($V$~$\sim$~11.2; Figure ~\ref{fig09} bottom panel).
Phase $\phi$~=~0.009 (repeated at $\phi$~=~1.009) corresponds to 
near-maximum positive (redward) skew, maximum outflow 
($\Delta$RV~$\simeq$~$-$32~\kmsec), and maximum brightness
($V$~$\sim$~10.3).

In both panels of Figure~\ref{fig08} the red curves are Gaussians 
fitted to the unskewed wings of the observed profiles.   
Gaussian $\sigma$ values of 12.8~\kmsec\ and 12.0~\kmsec\ were used to 
match these wings for phases 0.849 and 0.009, respectively.  
We assume that these red wings represent the intrinsic profiles of gas in 
the metal line-forming regions at these phases.
We then empirically added 4 additional line components with the same $\sigma$
values but weaker line depths and increasing RV offsets in order to adequately
match the skewed profile wings.
The observed profiles are formed in ranges of geometric depth that begin at 
the photosphere ($\tau$~$\sim$~0.6) and continue to very small values in 
the case of H$\alpha$. 
Most metal lines are formed at modest optical depths 
0.1~$\lesssim$~$\tau$~$\lesssim$~0.8 (\eg, \citealt{gray08}), with the far
wings arising from the deepest layers and cores at higher atmospheric optical
levels.
As lines grow stronger their mean formation level advances outward in 
stellar atmospheres.
In early stellar atmospheres modeling H$\alpha$ formation was considered
as arising in a separate outer atmosphere slab by \cite{kraft64}.
This notion is supported by the computations of \cite{stellingwerf13} that 
produced an accumulation of gas high in the atmosphere due to the succession 
of a multitude of outward progressing shocks.

\begin{figure}
\epsscale{0.60}
\plotone{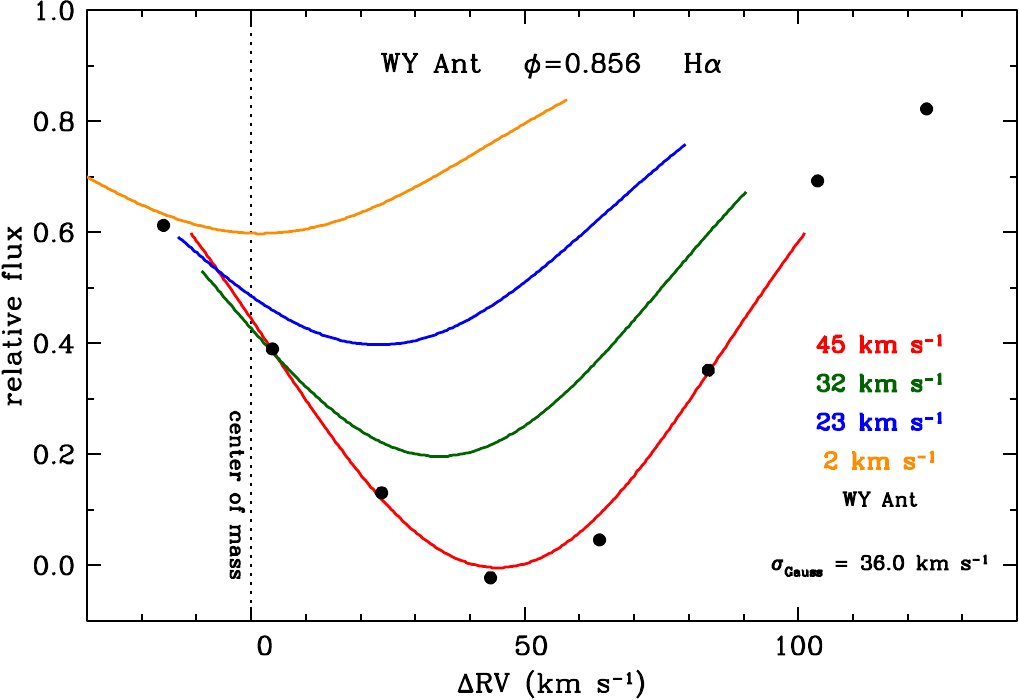}
\caption{\label{fig11}
\footnotesize
   H$\alpha$ profile for WY Ant 
   at phase $\phi$~=~0.86.
   The lines and symbols are as in Figure~\ref{fig08}
}
\end{figure}

To understand our skew profiles during rising light 
($\phi$~$\gtrsim$~0.9) we consider a simple model in which an outward 
propagating pressure wave, conserving momentum ($\rho v$ = constant), moves 
through an atmosphere of exponentially decreasing density.  
Conservation of momentum then puts a condition on the velocity of the wave, 
$v$ = cR$^{\alpha}$ in which $c$ and $\alpha$ are positive constants 
chosen to match the skew contributions at locations given by equations 
24 and 25 of \cite{edmonds69}.
In the post-maximum light phase ($\phi$~$\gtrsim$~0.1)
the velocity field, produced initially by gravitational infall, is modified 
by deceleration due to the pressure support predicted by \cite{hill72}.

We have made numerical experiments to estimate atmospheric line 
formation heights.
We adopted the static model atmospheres of \cite{kurucz11,kurucz18}\footnote{
http://kurucz.harvard.edu/linelists.html} 
for these calculations.
For phase $\phi$~=~0.847 our estimate of height above the photosphere for 
the model that we adopt for minimum light is shown in Figure~\ref{fig10}.
Unfortunately a similar plot for $\phi$~=~ 0.009 cannot be constructed from the 
observed line profiles of H$\alpha$ because of the major interfering 
contribution of the red-shifted component of the doubled line profile at 
this phase (see the H$\alpha$ profiles for our program star metal-poor RV~Oct
in Figure 18 of \cite{chadid17}.

In our model the observed metal line profile (Figure~\ref{fig08}
black dots) is created by contributions from a succession of layers at 
geometric depths within the line forming regions ($\Delta$R in 
Figure~\ref{fig10}) that move at different redial velocities indicated 
in the legends.  
We use observed $\Delta$RV values of Figure~\ref{fig08} as proxies for 
unknown heights, $\Delta$R, within line forming regions. 
Actual contribution functions must be calculated by use of equations 24 
and 25 of \cite{edmonds69}
We make no attempt to define these geometric heights in this 
initial reconnaissance.

Our profile in Figure~\ref{fig08} (top) requires a velocity gradient 
$\sim$20~km~sec$^{-2}$;
the range of $\Delta$R corresponding to the range 
0.67~$>$~$\tau$~$>$~0.1 in which metal lines are formed is some three times 
smaller than the total velocity gradient 
$\sim$60~km~sec$^{-2}$
which we deem appropriate for H$\alpha$ in Figure~\ref{fig11}.
The two observed profiles 
conform to expectations.
The curves
with minima at fluxes 0,2, 0,4, 0.6, and 0.8, fitted to the 
skewed wings, are idealized Gaussian contributions to the observed skews 
at geometric depths defined by equations 24 and 25 of \cite{edmonds69}.
We make no attempt to define these geometric depths in this 
first reconnaissance.

\subsection{Comparison of Metal and H$\alpha$ gradients}\label{gradcomp}

Figure~\ref{fig11} shows the H$\alpha$ profile in this star
at phase $\phi$~=~0.86, almost identical the 0.849 phase chosen for one
of the metal line plots in Figure~\ref{fig08}.\footnote{
The top panel of Figure~\ref{fig09} contains two sets of RV
measurements for H$\alpha$ in WY~Ant in the phase range
$\phi$~$\sim$~0.95$-$1.05.
This is the phase regime with obvious doubling of H$\alpha$ profiles
mentioned earlier, limiting our ability to make detailed statements about
this transition in our RRab stars.}
Comparison of the FWHMs of metal lines (12.0$-$12.8~\kmsec) and H$\alpha$
immediately shows that temperature, not turbulence, must be the principal 
cause of observed line broadening in our spectra.
Were turbulence the cause of broadening, line widths would be independent of 
atomic mass. 
To estimate temperatures we first remove instrumental broadening 
FWHM$_{inst}$ = 11.36~$\kmsec$ \citep{preston19} from the observed values.  
In the absence of better information about the temperature structure of 
the line forming regions we calculate Local Thermodynamic Equilibrium (LTE)
temperatures given in Table~\ref{tab-ltetemp}. 
We regard the factor of $\sim$2 disparity between our temperature estimates 
for the metal lines at phases 0.849 and 0.009 as an indicator of the 
uncertainty in our calculations.
We are content to conclude that temperatures of order 100,000~K in shocked 
gas are required to explain line widths in our spectra of WY~Ant, an RRab star 
typical of our metal-poor sample.

\section{SUMMARY}\label{summary}

We have investigated metal and H$\alpha$ line profile shapes in 17 
RRab and 5 RRc variable stars.
To create high resolution, high signal-to-noise line profiles for each
star we first performed coaddions of individual spectra in narrow phase 
intervals.
Then for each coadded spectrum we stacked together many metal lines in 
velocity space.
The resulting mean line profiles exhibit line profile distortions that vary
regularly with phase for all of our RRab stars.
No major differences in the line profiles variations are seen over a large
metallicity range.
The RRc stars show no obvious phase-based variations.
We have constructed model line profiles that suggest that velocity gradients 
in RR Lyrae photospheres can successfully account for the observed cyclic
RV variations.

\begin{acknowledgments}

We thank Giuseppe Bono, Vittorio Braga, 
Valentina D'Orazi and our referee 
for comments that have improved this paper.
As always we appreciate the efforts of the Carnegie support staff for their 
efficient, cheerful assistance in our work at LCO. 
NOIRLab IRAF is distributed by the Community Science and Data Center at NSF's 
NOIRLab, which is managed by the Association of Universities for Research 
in Astronomy (AURA) under a cooperative agreement with the National Science 
Foundation.
This work has been supported by NSF grant AST-1616040 (C.S.).

\end{acknowledgments}

\facilities{Du Pont (echelle spectrograph)}

\software{IRAF (Tody 1986, Tody 1993),
SPECTRE (Fitzpatrick \& Sneden 1987)}

\clearpage
\begin{center}
\begin{deluxetable}{lrrrrrrrrr}
\tabletypesize{\footnotesize}
\tablewidth{0pt}
\tablecaption{Program Star Data\label{tab-stars}}
\tablecolumns{10}
\tablehead{
\colhead{Name}                     &
\colhead{$P$}                      &
\colhead{$T_0$\tablenotemark{a}}   &
\colhead{$V$}                      &
\colhead{$V_{amp}$}                &
\colhead{[Fe/H]}                   &
\colhead{N$_{spec}$}               &
\colhead{N$_{spec}$}               &
\colhead{\#lines\tablenotemark{b}} &
\colhead{\#lines\tablenotemark{b}} \\
\colhead{}                         &
\colhead{(d)}                      &
\colhead{(d)}                      &
\colhead{(mag)}                    &
\colhead{(mag)}                    &
\colhead{}                         &
\colhead{raw}                      &
\colhead{coadded}                  &
\colhead{min}                      &
\colhead{max}                       
}
\startdata
\multicolumn{10}{c}{RRab Stars} \\
WY Ant        & 0.574344 & 3835.592 & 10.4 & 0.85 & $-$1.66 & 145 & 21 & 32 & 76 \\
SW Aqr        & 0.459303 & 1876.138 & 10.6 & 1.28 & $-$1.24 & 135 & 18 &  8 & 55 \\
XZ Aps        & 0.587266 & 3836.276 & 11.9 & 1.10 & $-$1.65 & 289 & 26 & 17 & 77 \\
X Ari         & 0.651170 & 1890.064 &  9.2 & 0.94 & $-$2.50 & 195 & 19 & 12 & 49 \\
RR Cet        & 0.553029 & 1900.312 &  9.3 & 0.82 & $-$1.52 &  96 & 20 & 19 & 65 \\
W Crt         & 0.412013 & 1871.640 & 10.9 & 1.10 & $-$0.76 & 170 & 23 & 16 & 65 \\
SX For        & 0.605342 & 1870.406 & 10.9 & 0.64 & $-$1.62 &  86 & 14 & 28 & 62 \\
VX Her        & 0.455359 & 2699.996 & 10.1 & 1.25 & $-$1.60 &  81 & 13 &  7 & 56 \\
V Ind         & 0.479602 & 1873.466 &  9.6 & 1.06 & $-$1.62 & 246 & 24 & 25 & 71 \\
SS Leo        & 0.626335 & 1873.056 & 10.5 & 1.00 & $-$1.88 & 110 & 21 & 17 & 73 \\
ST Leo        & 0.477984 & 5322.597 & 11.0 & 1.19 & $-$1.29 &  65 & 13 & 10 & 64 \\
RV Oct        & 0.571170 & 3835.895 & 10.5 & 1.13 & $-$1.34 & 227 & 26 & 19 & 69 \\
AV Peg        & 0.390382 & 5360.677 & 10.0 & 0.96 & $-$0.14 & 164 & 15 & 13 & 40 \\
HH Pup        & 0.390745 & 1869.661 & 10.6 & 1.24 & $-$0.95 & 175 & 24 & 31 & 74 \\
W Tuc         & 0.642243 & 5454.570 & 10.9 & 1.11 & $-$1.64 & 142 & 23 & 12 & 48 \\
CD Vel        & 0.573509 & 3835.914 & 11.7 & 0.87 & $-$1.71 & 208 & 20 & 14 & 64 \\
UU Vir        & 0.475609 & 1886.488 & 10.1 & 1.08 & $-$0.93 & 130 & 18 & 12 & 68 \\
\multicolumn{10}{c}{RRc Stars\tablenotemark{c}} \\
AS023706–4257.8 & 0.311326 & 1868.717 &  8.8 & 0.50 & $-$1.88 &  96 & 10 & 35 & 58 \\
AS081933–2358.2 & 0.285667 & 4900.180 & 10.4 & 0.28 & $-$2.50 &   6 &  3 &  8 & 17 \\
AS085254–0300.3 & 0.266902 & 4900.400 & 12.5 & 0.47 & $-$1.50 &   6 &  2 & 28 & 38 \\
AS211933–1507.0 & 0.273460 & 1873.307 & 11.1 & 0.50 & $-$1.50 &  30 &  8 & 46 & 56 \\
AS230659–4354.6 & 0.281130 & 5014.925 & 12.8 & 0.30 & $-$1.24 &  30 &  7 & 13 & 26 \\
\enddata

\tablenotetext{a}{$T_0$ = HJB $-$ 2,450,000.}
\tablenotetext{b}{\#lines(min) and \#lines(max) are the smallest and largest
                  number of lines contributing to the mean line profiles at
                  a single phase point for a star.}
\tablenotetext{c}{Names for RRc stars are from the All Sky Automated Survey 
                 (ASAS), \cite{pojmanski05} and references therein.}
\end{deluxetable}
\end{center}

\begin{center}
\begin{deluxetable}{lrrrrrrrr}
\tabletypesize{\footnotesize}
\tablewidth{0pt}
\tablecaption{Observed Line Broadening Parameters\label{tab-broaden}}
\tablecolumns{9}
\tablehead{
\colhead{Star}                     &
\colhead{phase}                    &
\colhead{\# lines}                 &
\colhead{V$_{B0.50}$}              &
\colhead{V$_{R0.50}$}              &
\colhead{V$_{B0.85}$}              &
\colhead{V$_{R0.85}$}              &
\colhead{$\bar{\mu}_{3,0.50}$}     &
\colhead{$\bar{\mu}_{3,0.85}$}     \\
\colhead{}                         &
\colhead{$\phi$}                   &
\colhead{}                         &
\colhead{\kmsec}                   &
\colhead{\kmsec}                   &
\colhead{\kmsec}                   &
\colhead{\kmsec}                   &
\colhead{}                         &
\colhead{}
}
\startdata
WY Ant & 0.009 & 41 & 12.3 & 15.1 & 20.1 & 26.4 &    0.13 &    0.17 \\
WY Ant & 0.057 & 46 & 11.8 & 14.3 & 20.0 & 24.5 &    0.11 &    0.12 \\
WY Ant & 0.085 & 46 & 11.9 & 13.5 & 20.7 & 23.6 &    0.08 &    0.09 \\
WY Ant & 0.159 & 65 & 11.0 & 11.8 & 19.1 & 19.6 &    0.05 &    0.03 \\
WY Ant & 0.261 & 71 &  9.6 &  9.8 & 16.4 & 16.3 &    0.03 &    0.01 \\
WY Ant & 0.346 & 76 &  9.1 &  9.3 & 15.8 & 15.5 &    0.02 & $-$0.01 \\
WY Ant & 0.450 & 72 &  9.9 &  9.9 & 17.2 & 16.3 &    0.01 & $-$0.03 \\
WY Ant & 0.547 & 77 & 11.9 & 10.9 & 20.8 & 17.8 & $-$0.02 & $-$0.07 \\
WY Ant & 0.658 & 71 & 15.1 & 12.7 & 24.9 & 20.2 & $-$0.04 & $-$0.10 \\
WY Ant & 0.749 & 68 & 14.4 & 12.9 & 23.4 & 20.6 & $-$0.02 & $-$0.06 \\
WY Ant & 0.849 & 65 & 15.0 & 13.4 & 25.2 & 21.5 & $-$0.02 & $-$0.08 \\
WY Ant & 0.881 & 43 & 17.0 & 14.0 & 27.3 & 22.8 & $-$0.06 & $-$0.07 \\
WY Ant & 0.905 & 35 & 16.3 & 18.0 & 28.4 & 30.4 &    0.06 &    0.03 \\
WY Ant & 0.904 & 27 & 15.5 & 17.1 & 25.6 & 33.9 &    0.09 &    0.18 \\
WY Ant & 0.915 & 35 & 12.6 & 15.3 & 21.7 & 28.3 &    0.12 &    0.15 \\
WY Ant & 0.927 & 39 & 12.2 & 14.0 & 20.8 & 25.0 &    0.09 &    0.11 \\
WY Ant & 0.939 & 34 & 12.6 & 13.9 & 20.7 & 25.7 &    0.08 &    0.13 \\
WY Ant & 0.953 & 34 & 12.3 & 15.2 & 20.6 & 27.1 &    0.14 &    0.15 \\
WY Ant & 0.964 & 32 & 12.5 & 16.2 & 20.1 & 28.4 &    0.15 &    0.18 \\
WY Ant & 0.973 & 36 & 11.1 & 14.9 & 20.1 & 34.6 &    0.24 &    0.29 \\
WY Ant & 0.987 & 35 & 12.2 & 16.0 & 20.1 & 27.1 &    0.16 &    0.19 \\
\enddata

\tablecomments{(This table is available in its entirety in 
machine-readable form.)}

\end{deluxetable}
\end{center}

\begin{center}
\begin{deluxetable}{ccccc}
\tabletypesize{\footnotesize}
\tablewidth{0pt}
\tablecaption{LTE Temperature Estimates\label{tab-ltetemp}}
\tablecolumns{5}
\tablehead{
\colhead{phase $\phi$}             &
\colhead{spectrum}                 &
\colhead{FWHM$_{obs}$}             &
\colhead{FWHM$_{corr}$}            &
\colhead{T$_{LTE}$}                \\
\colhead{}                         &
\colhead{}                         &
\colhead{($\kmsec$)}               &
\colhead{($\kmsec$)}               &
\colhead{(K)}
}
\startdata
0.849 &  metals    &  12.8 &   5.9 &  118000 \\
0.009 &  metals    &  12.2 &   4.5 &   67000 \\
0.856 &  H$\alpha$ &  36.0 &  34.2 &   70000 \\
\enddata

\end{deluxetable}
\end{center}


\clearpage
\begin{thebibliography}{}
\expandafter\ifx\csname natexlab\endcsname\relax\def\natexlab#1{#1}\fi
\providecommand{\url}[1]{\href{#1}{#1}}
\providecommand{\dodoi}[1]{doi:~\href{http://doi.org/#1}{\nolinkurl{#1}}}
\providecommand{\doeprint}[1]{\href{http://ascl.net/#1}{\nolinkurl{http://ascl.net/#1}}}
\providecommand{\doarXiv}[1]{\href{https://arxiv.org/abs/#1}{\nolinkurl{https://arxiv.org/abs/#1}}}

\bibitem[{{Bono} {et~al.}(1994){Bono}, {Caputo}, \& {Stellingwerf}}]{bono94}
{Bono}, G., {Caputo}, F., \& {Stellingwerf}, R.~F. 1994, ApJL, 432, L51,
  \dodoi{10.1086/187509}

\bibitem[{{Bono} {et~al.}(2000){Bono}, {Castellani}, \& {Marconi}}]{bono20b}
{Bono}, G., {Castellani}, V., \& {Marconi}, M. 2000, \apjl, 532, L129,
  \dodoi{10.1086/312582}

\bibitem[{{Bono} {et~al.}(2020){Bono}, {Braga}, {Crestani}, {Fabrizio},
  {Sneden}, {Marconi}, {Preston}, {Mullen}, {Gilligan}, {Fiorentino},
  {Pietrinferni}, {Altavilla}, {Buonanno}, {Chaboyer}, {da Silva}, {Dall'Ora},
  {Degl'Innocenti}, {Di Carlo}, {Ferraro}, {Grebel}, {Iannicola}, {Inno},
  {Kovtyukh}, {Kunder}, {Lemasle}, {Marengo}, {Marinoni}, {Marrese},
  {Mart{\'\i}nez-V{\'a}zquez}, {Matsunaga}, {Monelli}, {Neeley}, {Nonino},
  {Moroni}, {Prudil}, {Stetson}, {Th{\'e}venin}, {Tognelli}, {Valenti}, \&
  {Walker}}]{bono20}
{Bono}, G., {Braga}, V.~F., {Crestani}, J., {et~al.} 2020, \apjl, 896, L15,
  \dodoi{10.3847/2041-8213/ab9538}

\bibitem[{{Braga} {et~al.}(2021{\natexlab{a}}){Braga}, {Crestani}, {Fabrizio},
  {Bono}, {Sneden}, {Preston}, {Storm}, {Kamann}, {Latour}, {Lala}, {Lemasle},
  {Prudil}, {Altavilla}, {Chaboyer}, {Dall'Ora}, {Ferraro}, {Gilligan},
  {Fiorentino}, {Iannicola}, {Inno}, {Kwak}, {Marengo}, {Marinoni}, {Marrese},
  {Mart{\'\i}nez-V{\'a}zquez}, {Monelli}, {Mullen}, {Matsunaga}, {Neeley},
  {Stetson}, {Valenti}, \& {Zoccali}}]{fabrizio21}
{Braga}, V.~F., {Crestani}, J., {Fabrizio}, M., {et~al.} 2021{\natexlab{a}},
  \apj, 919, 85, \dodoi{10.3847/1538-4357/ac1074}

\bibitem[{{Braga} {et~al.}(2021{\natexlab{b}}){Braga}, {Crestani}, {Fabrizio},
  {Bono}, {Sneden}, {Preston}, {Storm}, {Kamann}, {Latour}, {Lala}, {Lemasle},
  {Prudil}, {Altavilla}, {Chaboyer}, {Dall'Ora}, {Ferraro}, {Gilligan},
  {Fiorentino}, {Iannicola}, {Inno}, {Kwak}, {Marengo}, {Marinoni}, {Marrese},
  {Mart{\'\i}nez-V{\'a}zquez}, {Monelli}, {Mullen}, {Matsunaga}, {Neeley},
  {Stetson}, {Valenti}, \& {Zoccali}}]{braga21}
---. 2021{\natexlab{b}}, \apj, 919, 85, \dodoi{10.3847/1538-4357/ac1074}

\bibitem[{{Chadid} {et~al.}(2017){Chadid}, {Sneden}, \& {Preston}}]{chadid17}
{Chadid}, M., {Sneden}, C., \& {Preston}, G.~W. 2017, ApJ, 835, 187,
  \dodoi{10.3847/1538-4357/835/2/187}

\bibitem[{{Crestani} {et~al.}(2021{\natexlab{a}}){Crestani}, {Fabrizio},
  {Braga}, {Sneden}, {Preston}, {Ferraro}, {Iannicola}, {Bono}, {Alves-Brito},
  {Nonino}, {D'Orazi}, {Inno}, {Monelli}, {Storm}, {Altavilla}, {Chaboyer},
  {Dall'Ora}, {Fiorentino}, {Gilligan}, {Grebel}, {Lala}, {Lemasle}, {Marengo},
  {Marinoni}, {Marrese}, {Mart{\'\i}nez-V{\'a}zquez}, {Matsunaga}, {Mullen},
  {Neeley}, {Prudil}, {da Silva}, {Stetson}, {Th{\'e}venin}, {Valenti},
  {Walker}, \& {Zoccali}}]{crestani21a}
{Crestani}, J., {Fabrizio}, M., {Braga}, V.~F., {et~al.} 2021{\natexlab{a}},
  \apj, 908, 20, \dodoi{10.3847/1538-4357/abd183}

\bibitem[{{Crestani} {et~al.}(2021{\natexlab{b}}){Crestani}, {Braga},
  {Fabrizio}, {Bono}, {Sneden}, {Preston}, {Ferraro}, {Iannicola}, {Nonino},
  {Fiorentino}, {Th{\'e}venin}, {Lemasle}, {Prudil}, {Alves-Brito},
  {Altavilla}, {Chaboyer}, {Dall'Ora}, {D'Orazi}, {Gilligan}, {Grebel},
  {Koch-Hansen}, {Lala}, {Marengo}, {Marinoni}, {Marrese},
  {Mart{\'\i}nez-V{\'a}zquez}, {Matsunaga}, {Monelli}, {Mullen}, {Neeley}, {da
  Silva}, {Stetson}, {Salaris}, {Storm}, {Valenti}, \& {Zoccali}}]{crestani21b}
{Crestani}, J., {Braga}, V.~F., {Fabrizio}, M., {et~al.} 2021{\natexlab{b}},
  \apj, 914, 10, \dodoi{10.3847/1538-4357/abfa23}

\bibitem[{{Dravins}(1982)}]{dravins82}
{Dravins}, D. 1982, \araa, 20, 61, \dodoi{10.1146/annurev.aa.20.090182.000425}

\bibitem[{{Edmonds}(1969)}]{edmonds69}
{Edmonds}, F.~N., J. 1969, \jqsrt, 9, 1427,
  \dodoi{10.1016/0022-4073(69)90126-5}

\bibitem[{{Fabrizio} {et~al.}(2019){Fabrizio}, {Bono}, {Braga}, {Magurno},
  {Marinoni}, {Marrese}, {Ferraro}, {Fiorentino}, {Giuffrida}, {Iannicola},
  {Monelli}, {Altavilla}, {Chaboyer}, {Dall'Ora}, {Gilligan}, {Layden},
  {Marengo}, {Nonino}, {Preston}, {Sesar}, {Sneden}, {Valenti}, {Th{\'e}venin},
  \& {Zoccali}}]{fabrizio19}
{Fabrizio}, M., {Bono}, G., {Braga}, V.~F., {et~al.} 2019, \apj, 882, 169,
  \dodoi{10.3847/1538-4357/ab3977}

\bibitem[{{Fitzpatrick} {et~al.}(2024){Fitzpatrick}, {Placco}, {Bolton},
  {Merino}, {Ridgway}, \& {Stanghellini}}]{fitzpatrick24}
{Fitzpatrick}, M., {Placco}, V., {Bolton}, A., {et~al.} 2024, arXiv e-prints,
  arXiv:2401.01982, \dodoi{10.48550/arXiv.2401.01982}

\bibitem[{{Fitzpatrick} \& {Sneden}(1987)}]{fitzpatrick87}
{Fitzpatrick}, M.~J., \& {Sneden}, C. 1987, in Bulletin of the American
  Astronomical Society, Vol.~19, Bulletin of the American Astronomical Society,
  1129

\bibitem[{{For} {et~al.}(2011{\natexlab{a}}){For}, {Preston}, \&
  {Sneden}}]{for11a}
{For}, B.-Q., {Preston}, G.~W., \& {Sneden}, C. 2011{\natexlab{a}}, ApJS, 194,
  38, \dodoi{10.1088/0067-0049/194/2/38}

\bibitem[{{For} {et~al.}(2011{\natexlab{b}}){For}, {Sneden}, \&
  {Preston}}]{for11b}
{For}, B.-Q., {Sneden}, C., \& {Preston}, G.~W. 2011{\natexlab{b}}, ApJS, 197,
  29, \dodoi{10.1088/0067-0049/197/2/29}

\bibitem[{{Govea} {et~al.}(2014){Govea}, {Gomez}, {Preston}, \&
  {Sneden}}]{govea14}
{Govea}, J., {Gomez}, T., {Preston}, G.~W., \& {Sneden}, C. 2014, ApJ, 782, 59,
  \dodoi{10.1088/0004-637X/782/2/59}

\bibitem[{{Gray}(2008)}]{gray08}
{Gray}, D.~F. 2008, {The Observation and Analysis of Stellar Photospheres}
  ({Cambridge, UK: Cambridge University Press})

\bibitem[{{Hill}(1972)}]{hill72}
{Hill}, S.~J. 1972, ApJ, 178, 793, \dodoi{10.1086/151835}

\bibitem[{{Kraft} {et~al.}(1964){Kraft}, {Preston}, \& {Wolff}}]{kraft64}
{Kraft}, R.~P., {Preston}, G.~W., \& {Wolff}, S.~C. 1964, \apj, 140, 235,
  \dodoi{10.1086/147910}

\bibitem[{{Kurucz}(2011)}]{kurucz11}
{Kurucz}, R.~L. 2011, Canadian Journal of Physics, 89, 417,
  \dodoi{10.1139/p10-104}

\bibitem[{{Kurucz}(2018)}]{kurucz18}
{Kurucz}, R.~L. 2018, in Astronomical Society of the Pacific Conference Series,
  Vol. 515, {Workshop on Astrophysical Opacities}, ed. C.~{Mendoza},
  S.~{Turck-Chi{\' e}ze}, \& J.~{Colgan}, 47

\bibitem[{{Moore} {et~al.}(1966){Moore}, {Minnaert}, \& {Houtgast}}]{moore66}
{Moore}, C.~E., {Minnaert}, M.~G.~J., \& {Houtgast}, J. 1966, {The solar
  spectrum 2935 A to 8770 A} (National Bureau of Standards Monograph,
  Washington: US Government Printing Office (USGPO))

\bibitem[{{Pojmanski} {et~al.}(2005){Pojmanski}, {Pilecki}, \&
  {Szczygiel}}]{pojmanski05}
{Pojmanski}, G., {Pilecki}, B., \& {Szczygiel}, D. 2005, \actaa, 55, 275

\bibitem[{{Preston} {et~al.}(2022){Preston}, {Sneden}, \& {Chadid}}]{preston22}
{Preston}, G.~W., {Sneden}, C., \& {Chadid}, M. 2022, \aj, 163, 109,
  \dodoi{10.3847/1538-3881/ac46ca}

\bibitem[{{Preston} {et~al.}(2019){Preston}, {Sneden}, {Chadid}, {Thompson}, \&
  {Shectman}}]{preston19}
{Preston}, G.~W., {Sneden}, C., {Chadid}, M., {Thompson}, I.~B., \& {Shectman},
  S.~A. 2019, \aj, 157, 153, \dodoi{10.3847/1538-3881/ab0ae1}

\bibitem[{{Sneden} {et~al.}(2017){Sneden}, {Preston}, {Chadid}, \&
  {Adam{\'o}w}}]{sneden17}
{Sneden}, C., {Preston}, G.~W., {Chadid}, M., \& {Adam{\'o}w}, M. 2017, ApJ,
  848, 68, \dodoi{10.3847/1538-4357/aa8b10}

\bibitem[{{Sneden} {et~al.}(2018){Sneden}, {Preston}, {Kollmeier}, {Crane},
  {Morrell}, {Prieto}, {Shectman}, {Skowron}, \& {Thompson}}]{sneden18}
{Sneden}, C., {Preston}, G.~W., {Kollmeier}, J.~A., {et~al.} 2018, AJ, 155, 45,
  \dodoi{10.3847/1538-3881/aa9f16}

\bibitem[{Sneden {et~al.}(2021)Sneden, Preston, Chadid, Shectman, \&
  Thompson}]{sneden21b}
Sneden, C., Preston, G.~W., Chadid, M., Shectman, S., \& Thompson, I. 2021,
  Complete RR Lyrae Las Campanas spectra,  Zenodo,
  \dodoi{10.5281/ZENODO.5794389}

\bibitem[{{Stellingwerf}(2013)}]{stellingwerf13}
{Stellingwerf}, R.~F. 2013, ArXiv e-prints.
\newblock \doarXiv{1310.0535}

\bibitem[{{Tody}(1986)}]{tody86}
{Tody}, D. 1986, in Society of Photo-Optical Instrumentation Engineers (SPIE)
  Conference Series, Vol. 627, Instrumentation in astronomy VI, ed. D.~L.
  {Crawford}, 733, \dodoi{10.1117/12.968154}

\bibitem[{{Tody}(1993)}]{tody93}
{Tody}, D. 1993, in Astronomical Society of the Pacific Conference Series,
  Vol.~52, Astronomical Data Analysis Software and Systems II, ed. R.~J.
  {Hanisch}, R.~J.~V. {Brissenden}, \& J.~{Barnes}, 173

\end{thebibliography}
\end{document}